\documentclass[superscriptaddress,prc,aps,twocolumn,showpacs]{revtex4}
\usepackage{amsmath,amssymb}
\usepackage{graphicx}
\usepackage{hyperref}
\newcommand{\Mainz}[1]
{\affiliation{Institut f\"ur Kernphysik, University of Mainz, D-55099 Mainz,Germany}}
\newcommand{\Bonn}[1]
{\affiliation{Helmholtz-Institut f\"ur Strahlen- und Kernphysik, University of Bonn,
 D-53115 Bonn, Germany}}
\newcommand{\Regina}[1]
{\affiliation{University of Regina, Regina, Saskatchewan S4S 0A2, Canada}}
\newcommand{\Kent}[1]
{\affiliation{Kent State University, Kent, Ohio 44242-0001, USA}}
\newcommand{\Glasgow}[1]
{\affiliation{SUPA School of Physics and Astronomy, University of Glasgow,
 Glasgow G12 8QQ, United Kingdom}}
\newcommand{\Giessen}[1]
{\affiliation{II Physikalisches Institut, University of Giessen, D-3539 Giessen, Germany}}
\newcommand{\Dubna}[1]
{\affiliation{Joint Institute for Nuclear Research, 141980 Dubna, Russia}}
\newcommand{\Pavia}[1]
{\affiliation{INFN Sezione di Pavia, I-27100 Pavia, Italy}}
\newcommand{\GWU}[1]
{\affiliation{The George Washington University, Washington, DC 20052-0001, USA}}
\newcommand{\LPI}[1]
{\affiliation{Lebedev Physical Institute, 119991 Moscow, Russia}}
\newcommand{\Dalhousie}[1]
{\affiliation{Dalhousie University, Halifax, Nova Scotia B3H 4R2, Canada}}
\newcommand{\Halifax}[1]
{\affiliation{Saint Mary’s University, Halifax, Nova Scotia B3H 3C3, Canada}}
\newcommand{\UniPavia}[1]
{\affiliation{Dipartimento di Fisica, Universit\`a di Pavia, I-27100 Pavia, Italy}}
\newcommand{\Basel}[1]
{\affiliation{Institut f\"ur Physik, University of Basel, CH-4056 Basel, Switzerland}}
\newcommand{\Edinburgh}[1]
{\affiliation{School of Physics, University of Edinburgh, Edinburgh EH9 3JZ,
 United Kingdom}}
\newcommand{\INR}[1]
{\affiliation{Institute for Nuclear Research, 125047 Moscow, Russia}}
\newcommand{\Sackville}[1]
{\affiliation{Mount Allison University, Sackville, New Brunswick E4L 1E6, Canada}}
\newcommand{\Zagreb}[1]
{\affiliation{Rudjer Boskovic Institute, HR-10000 Zagreb, Croatia}}
\newcommand{\ITEP}[1]
{\affiliation{Institute for Theoretical and Experimental Physics,
 SRC Kurchatov Institute, Moscow, 117218 Russia}}
\newcommand{\Amherst}[1]
{\affiliation{University of Massachusetts, Amherst, Massachusetts 01003, USA}}
\newcommand{\UCLA}[1]
{\affiliation{University of California Los Angeles, Los Angeles, California 90095-1547, USA}}
\newcommand{\Jerusalem}[1]
{\affiliation{Racah Institute of Physics, Hebrew University of Jerusalem, Jerusalem 91904, Israel}}
\begin{document}
\title{
Measurement of the $\pi^0 \to e^+e^-\gamma$ Dalitz decay at the Mainz Microtron}

\author{P.~Adlarson}\Mainz \\
\author{F.~Afzal}\Bonn \\
\author{P.~Aguar-Bartolom\'e}\Mainz \\
\author{Z.~Ahmed}\Regina \\
\author{C.~S.~Akondi}\Kent \\
\author{J.~R.~M.~Annand}\Glasgow \\
\author{H.~J.~Arends}\Mainz \\
\author{K.~Bantawa}\Kent \\
\author{R.~Beck}\Bonn \\
\author{H.~Bergh\"auser}\Giessen \\
\author{M.~Biroth}\Mainz \\
\author{N.~S.~Borisov}\Dubna \\
\author{A.~Braghieri}\Pavia \\
\author{W.~J.~Briscoe}\GWU \\
\author{S.~Cherepnya}\LPI \\
\author{F.~Cividini}\Mainz \\
\author{C.~Collicott}\Dalhousie \\ \Halifax \\
\author{S.~Costanza}\Pavia \\ \UniPavia \\
\author{A.~Denig}\Mainz \\
\author{M.~Dieterle}\Basel \\
\author{E.~J.~Downie}\Mainz \\ \GWU \\
\author{P.~Drexler}\Mainz \\
\author{M.~I.~Ferretti Bondy}\Mainz \\
\author{L.~V.~Fil'kov}\LPI \\
\author{S.~Gardner}\Glasgow \\
\author{S.~Garni}\Basel \\
\author{D.~I.~Glazier}\Glasgow \\ \Edinburgh \\
\author{D.~Glowa}\Edinburgh \\
\author{W.~Gradl}\Mainz \\
\author{G.~M.~Gurevich}\INR \\
\author{D.~J.~Hamilton}\Glasgow \\
\author{D.~Hornidge}\Sackville \\
\author{G.~M.~Huber}\Regina \\
\author{T.~C.~Jude}\Edinburgh \\
\author{A.~K\"aser}\Basel\\
\author{V.~L.~Kashevarov}\Mainz \\ \LPI \\
\author{S.~Kay}\Edinburgh \\
\author{I.~Keshelashvili}\Basel\\
\author{R.~Kondratiev}\INR \\
\author{M.~Korolija}\Zagreb \\
\author{B.~Krusche}\Basel \\
\author{V.~V.~Kulikov}\ITEP \\
\author{A.~Lazarev}\Dubna \\
\author{J.~Linturi}\Mainz \\
\author{V.~Lisin}\LPI \\
\author{K.~Livingston}\Glasgow \\
\author{I.~J.~D.~MacGregor}\Glasgow \\
\author{R.~Macrae}\Glasgow \\
\author{D.~M.~Manley}\Kent \\ 
\author{P.~P.~Martel}\Mainz \\ \Amherst \\
\author{M.~Martemianov}\ITEP \\
\author{J.~C.~McGeorge}\Glasgow \\
\author{E.~F.~McNicoll}\Glasgow \\
\author{V.~Metag}\Giessen \\
\author{D.~G.~Middleton}\Mainz \\ \Sackville \\
\author{R.~Miskimen}\Amherst \\
\author{E.~Mornacchi}\Mainz \\
\author{C.~Mullen}\Glasgow \\
\author{A.~Mushkarenkov}\Pavia \\ \Amherst \\ 
\author{A.~Neganov}\Dubna \\
\author{A.~Neiser}\Mainz \\ 
\author{A.~Nikolaev}\Bonn \\ 
\author{M.~Oberle}\Basel \\
\author{M.~Ostrick}\Mainz \\  
\author{P.~Ott}\Mainz \\   
\author{P.~B.~Otte}\Mainz \\
\author{D.~Paudyal}\Regina \\
\author{P.~Pedroni}\Pavia \\
\author{A.~Polonski}\INR \\  
\author{S.~Prakhov}\thanks{corresponding author, e-mail: prakhov@ucla.edu}\Mainz \\ \UCLA \\
\author{A.~Rajabi}\Amherst \\
\author{J.~Robinson}\Glasgow \\
\author{G.~Ron}\Jerusalem \\
\author{G.~Rosner}\Glasgow \\
\author{T.~Rostomyan}\Basel \\
\author{C.~Sfienti}\Mainz \\
\author{M.~H.~Sikora}\Edinburgh \\
\author{V.~Sokhoyan}\Mainz \\ \GWU \\
\author{K.~Spieker}\Bonn \\
\author{O.~Steffen}\Mainz \\
\author{I.~I.~Strakovsky}\GWU \\
\author{B.~Strandberg}\Glasgow \\
\author{Th.~Strub}\Basel \\
\author{I.~Supek}\Zagreb \\
\author{A.~Thiel}\Bonn \\
\author{M.~Thiel}\Mainz \\
\author{A.~Thomas}\Mainz \\   
\author{M.~Unverzagt}\Mainz \\ 
\author{Yu.~A.~Usov}\Dubna \\
\author{S.~Wagner}\Mainz \\
\author{N.~Walford}\Basel \\ 
\author{D.~P.~Watts}\Edinburgh \\
\author{D.~Werthm\"uller}\Glasgow \\ \Basel \\
\author{J.~Wettig}\Mainz \\
\author{L.~Witthauer}\Basel \\ 
\author{M.~Wolfes}\Mainz \\
\author{L.~A.~Zana}\Edinburgh \\

\collaboration{A2 Collaboration at MAMI}

\date{\today}
         
\begin{abstract}
 The Dalitz decay $\pi^0 \to e^+e^-\gamma$ has
 been measured in the $\gamma p\to \pi^0 p$ reaction
 with the A2 tagged-photon facility at the Mainz Microtron, MAMI.
 The value obtained for the slope parameter
 of the  $\pi^0$ electromagnetic transition form factor,
 $a_\pi=0.030\pm 0.010_{\mathrm{tot}}$,
 is in agreement with existing measurements of this decay
 and with recent theoretical calculations.
 The uncertainty obtained in the value of $a_\pi$
 is lower than in previous results based on
 the $\pi^0\to e^+e^-\gamma$ decay.
\end{abstract}

%\pacs{
% 14.40.Be, % Mesons - light mesons
% 13.20.-v, % Decay - mesons - leptonic and semileptonic
% 13.40.Gp  % Form Factors - electromagnetic
%}

\maketitle

\section{Introduction}

 The electromagnetic (e/m) transition form factors (TFFs) of light
 mesons play an important role in understanding the properties
 of these particles as well as in low-energy precision
 tests of the Standard Model (SM) and Quantum Chromodynamics (QCD)~\cite{TFFW12}.
 These TFFs appear as input information for data-driven approximations
 and model calculations, including such quantities as rare pseudoscalar
 decays~\cite{Leupold_2015,Pere_1512}.
 In particular, the TFFs of light mesons enter as contributions to the hadronic
 light-by-light (HLbL) scattering calculations~\cite{Colangelo_2014,Colangelo_2015}
 that are important for more accurate theoretical
 determinations of the anomalous magnetic moment of the muon, $(g-2)_\mu$,
 within the SM~\cite{g_2,Nyffeler_2016}.
 Recently, data-driven approaches, using dispersion relations, have been
 proposed~\cite{Colangelo_2014,Colangelo_2015,Pauk_2014} to attempt a better
 determination of the HLbL contribution to $(g-2)_\mu$ in a model-independent way.
 The precision of the calculations used to describe
 the HLbL contributions to $(g-2)_\mu$ can then be tested by directly comparing 
 theoretical predictions from these approaches for e/m TFFs
 of light mesons with experimental data.

 The TFF parameters that can be extracted from the Dalitz decay of the lightest meson,
 $\pi^0 \to e^+e^-\gamma$, are important to constrain calculations that estimate
 the pion-exchange term, $a_{\mu}^{\pi^0}$, to the HLbL scattering contribution
 to $(g-2)_\mu$~\cite{g_2}.
 The precise knowledge of the $\pi^0$ TFF is essential for a precision calculation of
 the decay width of the rare decay $\pi^0 \to e^+e^-$, the experimental value of which
 is in some disagreement with SM predictions ~\cite{Leupold_2015,Pere_1512}.
 In addition, this Dalitz decay recently attracted special attention because of
 a search for a hypothetical dark photon, $\gamma'$,
 that could be looked for here via the decay chain
 $\pi^0\to \gamma'\gamma \to e^+e^-\gamma$~\cite{Batell_2009,NA48_2_2015,WASA_COSY_2013}.
 
 For a structureless (pointlike) meson $A$, its decays into a lepton pair plus
 a photon, $A\to l^+l^-\gamma$, can be described within
 Quantum Electrodynamics (QED) via $A\to \gamma^*\gamma$, with the virtual photon $\gamma^*$
 decaying into the lepton pair~\cite{QED}. For the meson $A$, QED predicts a specific strong
 dependence of its decay rate on the dilepton invariant mass, $m_{ll}=q$.
 A deviation from the pure QED dependence, caused by the actual electromagnetic
 structure of the meson $A$, is formally described by its e/m TFF~\cite{Landsberg}.
 The Vector-Meson-Dominance (VMD) model~\cite{Sakurai} can be used to describe the coupling of
 the virtual photon $\gamma^*$ to the meson $A$ via an intermediate virtual vector meson $V$.
 This mechanism is especially strong in the timelike (the energy transfer larger than
 the momentum transfer) momentum-transfer region, $(2m_l)^2 < q^2 < m_A^2$, where
 a resonant behavior near $q^2 = m^2_V$ of the virtual photon arises because
 the virtual vector meson is approaching the mass shell~\cite{Landsberg},
 or even reaching it, as it is in the case of the $\eta'\to l^+l^-\gamma$ decay.    
 Experimentally, timelike TFFs can be determined by measuring the actual decay rate
 of $A\to l^+l^-\gamma$ as a function of the dilepton invariant mass $m_{ll}=q$,
 normalizing this dependence to the partial decay width $\Gamma(A\to \gamma\gamma)$,
 and then taking the ratio to the pure QED dependence for the decay rate of
 $A\to \gamma^*\gamma \to l^+l^-\gamma$.

 Because of the smallness of the $\pi^0$ mass, the virtual photon $\gamma^*$
 in the Dalitz decay of $\pi^0$
 can produce only the lightest lepton pair, $e^+e^-$, with $m_{ee}=q$.
 Based on QED, the decay rate of $\pi^0\to \gamma^*\gamma \to e^+e^-\gamma$ can
 be parametrized as~\cite{Landsberg}
\begin{eqnarray}
 & & \frac{d\Gamma(\pi^0\to e^+e^-\gamma)}{dm_{ee}\Gamma(\pi^0\to \gamma\gamma)} =
 \frac{4\alpha}{3\pi m_{ee}} \times
\nonumber 
\\ 
 & \times &  (1-\frac{4m^2_e}{m^2_{ee}})^{\frac{1}{2}}
  (1+\frac{2m^2_e}{m^2_{ee}}) (1-\frac{m^2_{ee}}{m^2_{\pi^0}})^{3}
  |F_{\pi^0\gamma}(m_{ee})|^2 =
\nonumber 
\\ 
 & = & [{\rm{QED}}(m_{ee})] |F_{\pi^0\gamma}(m_{ee})|^2,
\label{eqn:dgdm_pi0}
\end{eqnarray}
 where $F_{\pi^0\gamma}$ is the normalized TFF of the $\pi^0$ meson, $m_{\pi^0}$ and
 $m_{e}$ are the masses of the $\pi^0$ meson and $e^{+/-}$, respectively.
 Because of the smallness of the momentum-transfer range for the $\pi^0\to e^+e^-\gamma$
 decay, its normalized TFF is typically parametrized as~\cite{PDG}
\begin{equation}
 F_{\pi^0\gamma}(m_{ee}) = 1 + a_\pi\frac{m^2_{ee}}{m^2_{\pi^0}},
\label{eqn:Fm}
\end{equation}
 where the parameter $a_\pi$ reflects the TFF slope at $m_{ee}=0$.
 A simple VMD model incorporates only the $\rho$, $\omega$, and $\phi$
 resonances (in the narrow-width approximation) as virtual vector mesons
 driving the photon interaction in $A\to\gamma^*\gamma$.
 Using a quark model for the corresponding couplings leads to neglecting $\phi$
 and yields~\cite{Landsberg}
 $a_\pi/m^2_{\pi^0} = 0.5 (1 + m^2_{\rho}/m^2_{\omega})/m^2_{\rho}\approx 1.648$~GeV$^{-2}$ 
 (or $a_\pi \approx 0.0300$) for the $\pi^0$ Dalitz decay.
 A more modern VMD prediction, which also includes the $\phi$-meson contribution,
 leads to $a_\pi \approx 0.0305$~\cite{Hoferichter_2014}.

 Another feature of this decay amplitude is an angular anisotropy of the virtual
 photon decaying into the $e^+e^-$ pair, which also determines the density of
 events along $m^2(\gamma e^{+/-})$ of the $\pi^0\to e^+e^-\gamma$ Dalitz plot.
 For the $e^+$, $e^-$, and $\gamma$ in the $\pi^0$ rest frame,
 the angle $\theta^*$ between the direction of one of the leptons
 in the virtual-photon (or the dilepton) rest frame and the direction
 of the dilepton system (which is opposite to the $\gamma$ direction)
 follows the dependence~\cite{NA60_2016}
\begin{equation}
 f(\cos\theta^*) = 1 + \cos^2\theta^* + (\frac{2m_e}{m_{ee}})^2 \sin^2\theta^*,
\label{eqn:dtheta}
\end{equation}
 with the $\sin^2\theta^*$ term becoming very small when $m_{ee}\gg 2m_e$.

 Both the $[{\rm{QED}}(m_{ee})]$ term in Eq.~(\ref{eqn:dgdm_pi0}) and
 the angular dependence in Eq.~(\ref{eqn:dtheta}) represent only
 the leading-order term of the $\pi^0\to e^+e^-\gamma$ decay amplitude,
 and radiative corrections need to be considered for a more accurate
 calculation of $[{\rm{QED}}(m_{ee},\cos\theta^*)]$.
 The most recent calculations of radiative corrections to the differential
 decay rate of the Dalitz decay $\pi^0\to e^+e^-\gamma$ were reported
 in Ref.~\cite{Husek_2015}. In that paper, the results of the classical work
 of Mikaelian and Smith~\cite{MS_1972} were recalculated, and the missing
 one-photon irreducible contribution at the one-loop level was included.
 Typically radiative corrections make the angular dependence of
 the virtual-photon decay weaker. For the $\pi^0$ Dalitz decay,
 the corrected $[{\rm{QED}}]$ term integrated over $\cos\theta^*$ is $\sim$1\%
 larger than the leading-order term at $q=15$~MeV and becomes $\sim$10\% lower
 at $q=120$~MeV.
   
 Despite the existence of recent high-statistics experiments searching for
 a dark-photon signal in $\pi^0\to e^+e^-\gamma$ decays~\cite{NA48_2_2015,WASA_COSY_2013},
 the magnitude of the Dalitz-decay slope parameter $a_\pi$ and its uncertainty
 in the Review of Particle Physics (RPP)~\cite{PDG}, $a_\pi = 0.032\pm0.004$,
 are mostly determined by a measurement of the spacelike $\pi^0$ TFF in
 the process $e^+e^-\to e^+e^-\pi^0$ by the CELLO detector~\cite{CELLO_1991}.
 Extrapolating this spacelike TFF under the assumption of the validity of VMD,
 the value $a_\pi = 0.0326\pm0.0026_{\mathrm{stat}}\pm0.0026_{\mathrm{syst}}$
 has been extracted. It should be noted, however, that this result not only
 introduces a certain model dependence, but also requires an extrapolation
 from the range of momentum transfers ($q^2>0.5$~GeV$^2$),
 where the actual measurement took place, toward small $q^2$.
 Further improvement in measuring the spacelike $\pi^0$ TFF in
 the process $e^+e^-\to e^+e^-\pi^0$ is expected from the BESIII
 detector~\cite{BESIII_prcom}.
 Because this measurement will cover smaller $q^2$, the precision in
 the slope parameter obtained by the extrapolation could be improved even more.

 To check the consistency of the $a_\pi$ values extracted
 from measurements at negative and positive $q^2$,
 the precision in the slope parameter obtained from measuring
 the Dalitz decays should be comparable with the results
 of extrapolating the spacelike TFFs.
 So far, the most accurate slope-parameter value obtained from measuring
 $\pi^0\to e^+e^-\gamma$ decays,
 $a_\pi = 0.025\pm0.014_{\mathrm{stat}}\pm0.026_{\mathrm{syst}}$~\cite{SINDRUM_1992},
 has uncertainties one order of magnitude larger than the value from CELLO~\cite{CELLO_1991}.
 This timelike measurement is based on the analysis of just
 $54 \cdot 10^3$ $\pi^0\to e^+e^-\gamma$ decays, with radiative corrections
 according to Ref.~\cite{MS_1972}, and does not provide
 any $|F_{\pi^0\gamma}(m_{ee})|^2$ data points.
 The results of the present work are going to improve the experimental situation for
 the timelike $\pi^0$ TFF, with the experimental statistic of $\pi^0\to e^+e^-\gamma$ decays
 larger by one order of magnitude, compared to Ref.~\cite{SINDRUM_1992}. 
 Further improvement in the timelike region is expected to be made by the NA62
 experiment, the preliminary result of which,
 $a_\pi = 0.0370\pm0.0053_{\mathrm{stat}}\pm0.0036_{\mathrm{syst}}$,
 was based on $1.05 \cdot 10^6$ $\pi^0\to e^+e^-\gamma$ decays observed~\cite{NA62_2016_prem}.
 The latest NA62 value for the slope parameter, which appeared after this paper
 was submitted for publication, updated their result to
 $a_\pi = 0.0368\pm0.0051_{\mathrm{stat}}\pm0.0025_{\mathrm{syst}} = 0.0368\pm0.0057_{\mathrm{tot}}$,
 based on $1.11 \cdot 10^6$ $\pi^0\to e^+e^-\gamma$ decays observed~\cite{NA62_2016_final}.

 Recent theoretical calculations for the $\pi^0\to \gamma^*\gamma \to e^+e^-\gamma$ TFF,
 in addition to the slope parameter $a_\pi$, also involve the curvature
 parameter $b_\pi$:
\begin{equation}
 F_{\pi^0\gamma}(m_{ee}) = 1 + a_\pi\frac{m^2_{ee}}{m^2_{\pi^0}}+ b_\pi\frac{m^4_{ee}}{m^4_{\pi^0}}~.
\label{eqn:Fm2}
\end{equation}

 A calculation based on a model-independent method using Pad\'e approximants
 was reported in Ref.~\cite{Mas12}.
 The analysis of spacelike data (CELLO~\cite{CELLO_1991}, CLEO~\cite{CLEO_1998},
 BABAR~\cite{BABAR_2011}, and Belle~\cite{Belle_2012}) with this method provides
 a good and systematic description of the low energy region, resulting in
 $a_\pi = 0.0324\pm0.0012_{\mathrm{stat}}\pm0.0019_{\mathrm{syst}}$ and
 $b_\pi = (1.06\pm0.09_{\mathrm{stat}}\pm0.25_{\mathrm{syst}})\cdot 10^{-3}$.
 Values with even smaller uncertainties, $a_\pi = 0.0307\pm0.0006$ and
 $b_\pi = (1.10\pm0.02)\cdot 10^{-3}$, were recently obtained by using
 dispersion theory~\cite{Hoferichter_2014}.
 In that analysis, the singly virtual TFF was calculated in both
 the timelike and the spacelike regions, based
 on data for the $e^+e^-\to 3\pi$ cross section, generalizing previous
 studies on $\omega/\phi \to 3\pi$ decays~\cite{Niecknig_2012} and
 $\gamma\pi \to \pi\pi$ scattering~\cite{Hoferichter_2012},
 and verifying the results by comparing them to timelike $e^+e^-\to \pi^0\gamma$ data
 at larger momentum transfer.
  
 The capability of the A2 experimental setup to measure Dalitz decays
 was demonstrated in Refs.~\cite{eta_tff_a2_2014,eta_tff_a2_2011} for 
 $\eta \to e^+e^-\gamma$. Measuring $\pi^0\to e^+e^-\gamma$
 is challenging because of the smallness of the TFF effect in the region
 of very low momentum transfer; the magnitude of $|F_{\pi^0\gamma}|^2$ is
 expected to reach only a 5\% enhancement above the pure QED dependence 
 at $m_{ee}=120$~MeV/$c^2$. Thus, such a measurement requires
 high statistics to reach a statistical accuracy comparable with
 the expected TFF effect. Also, the magnitude of systematic uncertainties
 caused by the acceptance determination, background subtraction, and
 experimental resolutions needs to be small. 
 The advantage of measuring $\pi^0\to e^+e^-\gamma$ with the A2 setup
 at MAMI is that $\pi^0$ mesons can be produced in the reaction
 $\gamma p\to \pi^0 p$, which has a very large cross section at
 energies close to the $\Delta(1232)$ state, and there is no background
 from other physical reactions at these energies.
 The only background for $\pi^0\to e^+e^-\gamma$ decays
 are $\pi^0\to \gamma\gamma$ decays with a photon converting into an $e^+e^-$ pair
 in the material in front of electromagnetic calorimeters.
  
 New results for the $\pi^0\gamma$ e/m TFF
 presented in this paper are based on an analysis of $\sim4\cdot 10^5$
 $\pi^0 \to e^+e^-\gamma$ decays detected in the A2 experimental setup
 and using the radiative corrections from Ref.~\cite{Husek_2015}.
 In addition to a value for the slope parameter $a_\pi$,
 the present TFF results include $|F_{\pi^0\gamma}(m_{ee})|^2$ data points
 with their total uncertainties, which allows a more fair comparison of the data
 with theoretical calculations or the use of the data in model-independent fits.
 Previously, the same A2 data sets were used for measuring
 $\pi^0$ photoproduction on the proton~\cite{hornidge13,pi0_a2_2015}.

\section{Experimental setup}
\label{sec:Setup}

The process $\gamma p\to \pi^0 p \to e^+e^-\gamma p$
was measured by using the Crystal Ball (CB)~\cite{CB}
as a central calorimeter and TAPS~\cite{TAPS,TAPS2}
as a forward calorimeter. These detectors were
installed in the energy-tagged bremsstrahlung photon beam of
the Mainz Microtron (MAMI)~\cite{MAMI,MAMIC}. 
The photon energies were determined by using
the Glasgow--Mainz tagging spectrometer~\cite{TAGGER,TAGGER1,TAGGER2}.

The CB detector is a sphere consisting of 672
optically isolated NaI(Tl) crystals, shaped as
truncated triangular pyramids, which point toward
the center of the sphere. The crystals are arranged in two
hemispheres that cover 93\% of $4\pi$, sitting
outside a central spherical cavity with a radius of
25~cm, which holds the target and inner
detectors. In this experiment, TAPS was
arranged in a plane consisting of 384 BaF$_2$
counters of hexagonal cross section.
It was installed 1.5~m downstream of the CB center
and covered the full azimuthal range for polar angles
from $1^\circ$ to $20^\circ$.
More details on the energy and angular resolution of the CB and TAPS
are given in Refs.~\cite{slopemamic,etamamic}.

 The present measurement used electron beams
 with energies 855 and 1557 MeV from the Mainz Microtron, MAMI-C~\cite{MAMIC}.
 The data with the 855-MeV beam were taken in 2008 (Run-I)
 and those with the 1557-MeV beam in 2013 (Run-II).
 Bremsstrahlung photons, produced by the beam electrons
 in a radiator (100-$\mu$m-thick diamond and 10-$\mu$m Cu for Run-I and Run-II,
 respectively) and collimated by a Pb collimator (with diameter 3 and 4 mm
 for Run-I and Run-II, respectively),
 were incident on a 10-cm-long liquid hydrogen (LH$_2$) target located
 in the center of the CB.
 The total amount of material around the LH$_2$ target,
 including the Kapton cell and the 1-mm-thick carbon-fiber beamline,
 was equivalent to 0.8\% of a radiation length $X_0$.
 In the present measurement, it was essential to keep the material budget
 as low as possible to minimize the background from $\pi^0 \to \gamma\gamma$ decays
 with conversion of the photons into $e^+e^-$ pairs.
\begin{figure}
\includegraphics[width=7.cm,height=8.cm,bbllx=0.cm,bblly=0.cm,bburx=10.5cm,bbury=12.cm]{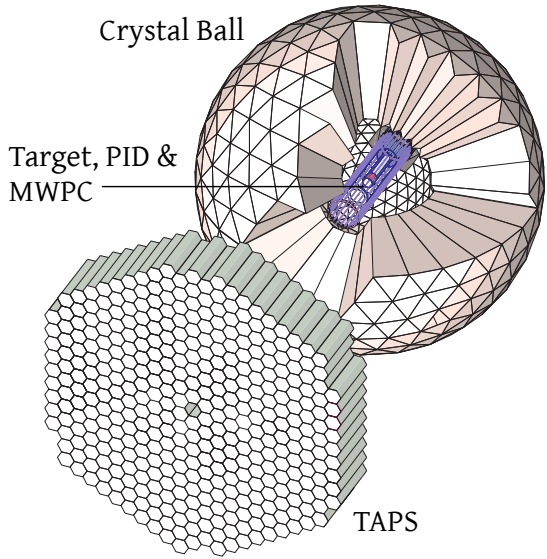}
\caption{(Color online)
  A general sketch of the Crystal Ball, TAPS, and particle identification (PID) detectors.
}
 \label{fig:cb_taps_pid} 
\end{figure}

 The target was surrounded by a Particle IDentification
 (PID) detector~\cite{PID} used to distinguish between charged and
 neutral particles. It is made of 24 scintillator bars
 (50 cm long, 4 mm thick) arranged as a cylinder with a radius of 12 cm.
 A general sketch of the CB, TAPS, and PID is shown
 in Fig.~\ref{fig:cb_taps_pid}.
 A multi-wire proportional chamber, MWPC, also shown in this figure
 (which consists of two cylindrical MWPCs inside each other),
 was not used in the present measurements because of its relatively low
 efficiency for detecting $e^{+/-}$.

 In Run-I, the energies of the incident photons were analyzed from
 140 up to 798~MeV by detecting the postbremsstrahlung electrons
 in the Glasgow tagged-photon spectrometer
 (Glasgow tagger)~\cite{TAGGER,TAGGER1,TAGGER2},
 and from 216 up to 1448~MeV in Run-II.
 The uncertainty in the energy of the tagged photons is mainly determined
 by the segmentation of tagger focal-plane detector in combination with
 the energy of the MAMI electron beam used in the experiments.
 Increasing the MAMI energy increases the energy range covered
 by the spectrometer and also has the corresponding effect on the uncertainty
 in $E_\gamma$. For the MAMI energy settings of 855 and 1557~MeV,
 this uncertainty was about $\pm 1$~MeV and $\pm 2$~MeV, respectively.
 More details on the tagger energy calibration and uncertainties
 in the energies can be found in Ref.~\cite{EtaMassA2}.

 The experimental trigger in Run-I required the total energy deposited in the CB
 to exceed $\sim$100~MeV and the number of so-called hardware clusters
 in the CB (multiplicity trigger) to be two or more.
 In the trigger, a hardware cluster in the CB was a block of 16
 adjacent crystals in which at least one crystal had an energy
 deposit larger than 30 MeV.
 In Run-II, the trigger only required the total energy
 in the CB to exceed $\sim$120~MeV.
 More details on the experimental conditions of Run-I and Run-II
 can be found in Refs.~\cite{hornidge13,pi0_a2_2015}.

\section{Data handling}
\label{sec:Data}
\subsection{Event selection}
\label{subsec:Data-I}

 To search for a signal from $\pi^0 \to e^+e^-\gamma$ decays, 
 candidates for the process $\gamma p\to e^+e^-\gamma p$
 were extracted from events having three or four clusters
 reconstructed by a software analysis in the CB and TAPS together.
 The offline cluster algorithm was optimized for finding
 a group of adjacent crystals in which the energy was deposited
 by a single-photon e/m shower. This algorithm works well for $e^{+/-}$,
 which also produce e/m showers in the CB and TAPS, and for proton clusters.
 The software threshold for the cluster energy was chosen to be 12 MeV.
 For the $\gamma p\to e^+e^-\gamma p$ candidates, 
 the three-cluster events were analyzed assuming that the final-state
 proton was not detected. To diminish possible background from
 $\gamma p\to \pi^0\pi^0p$ and $\gamma p\to \pi^0\pi^+n$, the selected
 energy range was limited to $E_{\gamma}<450$~MeV. To take the energies
 with the largest $\pi^0$ cross sections, $E_{\gamma}>167$~MeV was required
 for Run-I and $E_{\gamma}>216$~MeV for Run-II, in which the lower $E_{\gamma}$
 were not tagged. Note that a large fraction of $\pi^0$ events in this
 energy range are produced with the recoil proton below its detection threshold.

 The selection of candidate events and the reconstruction of the reaction
 kinematics were based on the kinematic-fit technique.
 Details of the kinematic-fit parametrization of the detector
 information and resolutions are given in Ref.~\cite{slopemamic}.
 Because the three-cluster sample, in which there are good
 $\gamma p\to \pi^0 p \to e^+e^-\gamma p$ events without the outgoing
 proton detected, was mostly dominated by $\gamma p\to \pi^0p\to \gamma\gamma p$
 events, the latter kinematic-fit hypothesis was tested first. Then all events for which
 the confidence level (CL) to be $\gamma p\to \pi^0p\to \gamma\gamma p$
 was greater than $10^{-5}$ were discarded from further analysis. It was
 checked that such a preselection practically does not cause any losses of
 $\pi^0 \to e^+e^-\gamma$ decays, but rejects a significant background
 from two-photon final states.
 Because e/m showers from electrons and positrons are
 very similar to those of photons, 
 the hypothesis $\gamma p \to 3\gamma p$ was tested to identify
 the $\gamma p\to e^+e^-\gamma p$ candidates.
 The events that satisfied this hypothesis with the CL greater
 than 1\% were accepted for further analysis. The kinematic-fit output
 was used to reconstruct the kinematics of the outgoing particles.
 In this output, there was no separation between e/m showers
 caused by the outgoing photon, electron, or positron.  
 Because the main purpose of the experiments was to measure
 the $\pi^0 \to e^+e^-\gamma$ decay rate
 as a function of the invariant mass $m(e^+e^-)$, the next step
 in the analysis was the separation of $e^+e^-$ pairs from
 final-state photons. This procedure was optimized by using a Monte
 Carlo (MC) simulation of the signal events.

 Because of the limited experimental resolution in the invariant mass $m(e^+e^-)$
 (the average value of $\sigma_m$ for which was $\sim$5.7 and $\sim$6.0~MeV
 for Run-I and Run-II, respectively) and the detection threshold for particles
 in the experimental setup, the MC simulation was made to be as similar as
 possible to the real $\gamma p\to \pi^0 p \to e^+e^-\gamma p$ events.
 This condition was important to minimize systematic uncertainties in the determination
 of experimental acceptances and to measure the TFF energy dependence properly.
 To reproduce the experimental yield of $\pi^0$ mesons and
 their angular distributions as a function of the incident-photon
 energy, the $\gamma p\to \pi^0 p$ reaction was generated according to
 the numbers of the corresponding $\pi^0$ events and their
 angular distributions measured in the same experiments~\cite{hornidge13,pi0_a2_2015}.
 The $\pi^0 \to e^+e^-\gamma$ decays were generated according to Eq.~(\ref{eqn:dgdm_pi0}),
 with the phase-space term removed and assuming the RPP value,
 $a_\pi = 0.032$~\cite{PDG}, for the TFF dependence.
 The angular dependence of the virtual photon decaying into the $e^+e^-$ pair
 was generated according to Eq.~(\ref{eqn:dtheta}).
 Then these dependences from the leading-order QED term of the decay amplitude
 were convoluted with radiative corrections based on the calculations
 of Ref.~\cite{Husek_2015}.
 The event vertices were generated uniformly along the 10-cm-long LH$_2$ target.

 The main background process, $\gamma p\to \pi^0p\to \gamma\gamma p$, was also studied
 by using the MC simulation. The yield and the production angular distributions of
 $\gamma p\to \pi^0p$ were generated in the same way as 
 for the process $\gamma p\to \pi^0 p \to e^+e^-\gamma p$. 

 For both $\pi^0$ decay modes, the generated events
 were propagated through a {\sc GEANT} (version 3.21) simulation of the experimental
 setup. To reproduce the resolutions observed in the experimental data, the {\sc GEANT}
 output (energy and timing) was subject to additional smearing, thus
 allowing both the simulated and experimental data to be analyzed in the same way.
 Matching the energy resolution between the experimental and MC events
 was achieved by adjusting the invariant-mass resolutions,
 the kinematic-fit stretch functions (or pulls), and probability
 distributions. Such an adjustment was based on the analysis of the
 same data sets for the reaction $\gamma p\to \pi^0 p\to \gamma\gamma p$,
 having almost no background from other physical reactions at these energies.
 The simulated events were also tested to check whether they passed
 the trigger requirements.
 
 The PID detector was used to identify the final-state $e^+e^-$ pair
 in the events initially selected as $\gamma p \to 3\gamma p$ candidates.
 Note that the detection efficiency for $e^{+/-}$ that pass through the PID is close
 to 100\%. Because, with respect to the LH$_2$ target, the PID provides a full
 coverage merely for the CB crystals, only events with three e/m showers in the CB
 were selected for further analysis. This criterion also made all selected events
 pass the trigger requirements on both the total energy in the CB (Run-I and Run-II)
 and the multiplicity (Run-I).
 The identification of $e^{+/-}$ in the CB was based on a correlation
 between the $\phi$ angles of fired PID elements with the angles
 of e/m showers in the calorimeter.
 The MC simulation of $\gamma p\to \pi^0 p \to e^+e^-\gamma p$ 
 was used to optimize this procedure, minimizing the probability for misidentification
 of $e^{+/-}$ with the final-state photons. This procedure was optimized with respect to
 how close an e/m shower in the CB should be to a fired PID element to be considered
 as $e^{+/-}$ (namely $\Delta\phi<18^\circ$), and how far it should be to be considered
 as a photon ($\Delta\phi>20^\circ$). This optimization decreases the efficiency
 in selecting true events for which the $\phi$ angle of the electron or the positron
 is close to the photon $\phi$ angle.
\begin{figure*}
\includegraphics[width=0.9\textwidth]{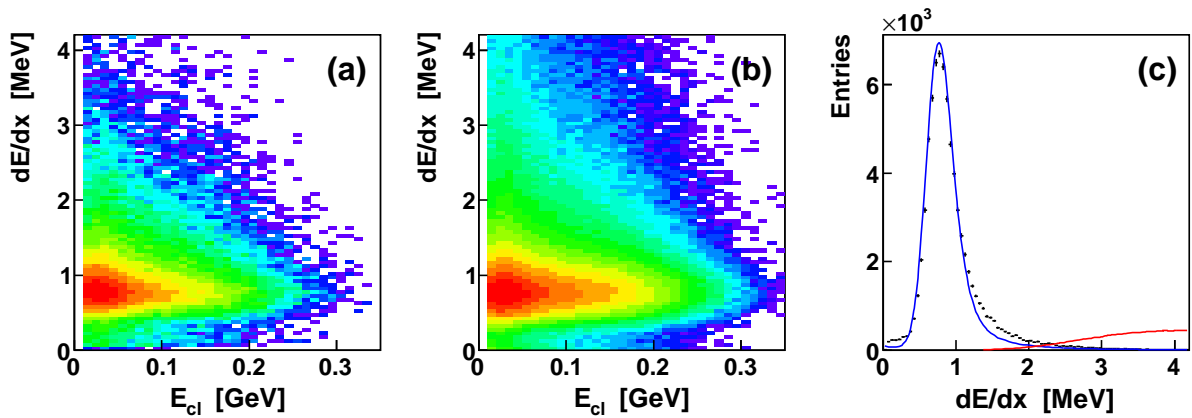}
\caption{ (Color online)
 Comparison of the $e^{+/-}$ $dE/dx$ of the PID for experimental
 $\pi^0 \to e^+e^-\gamma$ decays and the MC simulation.
 The two-dimensional density distribution (with logarithmic scale
 along plot axis $z$) for the $e^{+/-}$ $dE/dx$
 of the PID versus the energy of the corresponding clusters in the CB
 is shown in (a) for the experimental data of Run-I and in (b) for the MC simulation.
 The $e^{+/-}$ $dE/dx$ distributions for the experimental data (crosses) and
 the MC simulation (blue solid line) are compared in (c). The $dE/dx$ distribution
 from the recoil protons for the selected four-cluster events is shown in (c)
 by a red solid line. 
}
 \label{fig:pi0eeg_pid_dedx} 
\end{figure*}

 The analysis of the MC simulation for the main background reaction
 $\gamma p\to \pi^0 p \to \gamma\gamma p$ revealed that this process
 could mimic $\pi^0 \to e^+e^-\gamma$ events when one of the final-state photons
 converted into an $e^+e^-$ pair in the material between the production vertex and
 the NaI(Tl) surface. Because the opening angle between such electrons and positrons
 is typically very small, this background contributes mostly to low invariant
 masses $m(e^+e^-)$. A significant suppression of this background can be reached
 by requiring $e^+$ and $e^-$ to be identified by different PID elements.
 However, such a requirement also decreases the detection efficiency for
 actual $\pi^0 \to e^+e^-\gamma$ events, especially at low invariant masses $m(e^+e^-)$.
 In further analysis of $\pi^0 \to e^+e^-\gamma$ events, both options, with larger
 and smaller background remaining from $\pi^0\to\gamma\gamma$, were tested.

 Another background source from $\gamma p\to \pi^0 p \to \gamma\gamma p$ are
 events that survived the CL$<10^{-5}$ cut from testing this hypothesis itself.
 If one photon deposits some energy in the PID, then this e/m shower, together with
 the recoil proton, could be misidentified as an $e^+e^-$ pair. Such background does not
 mimic the $\pi^0 \to e^+e^-\gamma$ peak, but the suppression of this background
 improves the signal-to-background ratio, which is important for more
 reliable fitting of the signal peak above the remaining background.
 Similar background can come from the $\gamma p\to \pi^0 p \to e^+e^-\gamma p$
 events themselves when one of the leptons failed to be detected, and the recoil
 proton was misidentified with this lepton.
 The background from the misidentification of the recoil proton
 with $e^{+/-}$ can be suppressed by the analysis of energy losses,
 $dE/dx$, in the PID elements. To reflect the actual differential energy
 deposit $dE/dx$ in the PID, the energy signal from each element,
 ascribed to either $e^+$ or $e^-$, was multiplied by the sine
 of the polar angle of the corresponding particle,
 the magnitude of which was taken from the kinematic-fit output.
 All PID elements were calibrated so that the $e^{+/-}$ peak position matched
 the corresponding peak in the MC simulation.
 To reproduce the actual energy resolution of the PID with the MC simulation,  
 the {\sc GEANT} output for PID energies was subject to additional smearing,
 allowing the $e^{+/-}$ selection with $dE/dx$ cuts to be very similar for
 the experimental data and MC. The PID energy resolution in the MC simulations
 was adjusted to match the experimental $dE/dx$ spectra for the $e^{+/-}$ particles
 from $\pi^0 \to e^+e^-\gamma$ decays observed experimentally.
 Possible systematic uncertainties due to the $dE/dx$ cuts were checked
 via the stability of the results after narrowing the $dE/dx$ range for selecting $e^{+/-}$.

 The experimental $dE/dx$ resolution of the PID for $e^{+/-}$ in Run-I and the comparison
 of it with the MC simulation is illustrated in Fig.~\ref{fig:pi0eeg_pid_dedx}.
 Figures~\ref{fig:pi0eeg_pid_dedx}(a) and~(b) show 
 (for the experimental data and the MC simulation, respectively) two-dimensional plots
 of the $e^{+/-}$ $dE/dx$ value of the PID versus the energy of the corresponding
 clusters in the CB. As seen, there is no $dE/dx$ dependence of $e^{+/-}$
 on their energy in the CB, and applying cuts just on a $dE/dx$ value is sufficient
 for suppressing backgrounds caused by misidentifying protons as $e^{+/-}$.
 The comparison of the experimental $e^{+/-}$ $dE/dx$ distributions with
 the MC simulation is depicted in Fig.~\ref{fig:pi0eeg_pid_dedx}(c).
 A small difference in the tails of the $e^{+/-}$ peak can mostly be explained
 by some background remaining in the experimental spectrum. This background includes
 events with misidentified recoil protons, photons converting before
 reaching the crystal surface, and also a small fraction from accidental hits in the PID.
 The $dE/dx$ distribution from the recoil protons for the selected four-cluster events
 is shown in Fig.~\ref{fig:pi0eeg_pid_dedx}(c) by the red line, illustrating a quite small
 overlapping range of $e^{+/-}$ and the protons. Typical PID cuts, which were tested,
 varied from requiring $dE/dx<3.7$~MeV to $dE/dx<2.7$~MeV to suppress background events
 with misidentified protons, showing no systematic effects in the final results. 
 
 In addition to the background contributions discussed above,
 there are two more background sources.
 The first source comes from interactions of incident photons in the windows
 of the target cell. The subtraction of this background was based on the
 analysis of data samples that were taken with an empty target. The weight for
 the subtraction of the empty-target spectra was taken as a ratio of
 the photon-beam fluxes for the data samples with the full and the empty target.
 Another background was caused by random coincidences
 of the tagger counts with the experimental trigger;
 its subtraction was carried out by using 
 event samples for which all coincidences were random
 (see Refs.~\cite{slopemamic,etamamic} for more details).

\subsection{Analysis of $\pi^0 \to e^+e^-\gamma$ decays}
\label{subsec:Data-II}
 To measure the $\pi^0 \to e^+e^-\gamma$ yield as a function of
 the invariant mass $m(e^+e^-)$, the selected
 candidate events were divided into several $m(e^+e^-)$ bins.
 Events with $m(e^+e^-)<15$~MeV/$c^2$ were not analyzed at all,
 because e/m showers from those $e^+$ and $e^-$
 start to overlap too much in the CB.
 The number of $\pi^0 \to e^+e^-\gamma$ decays
 in every $m(e^+e^-)$ bin was determined by fitting
 the experimental $m(e^+e^-\gamma)$ spectra
 with the $\pi^0$ peak rising above a smooth background.
\begin{figure}
\includegraphics[width=0.45\textwidth,height=4.35cm,bbllx=1.cm,bblly=.5cm,bburx=19.5cm,bbury=9.cm]{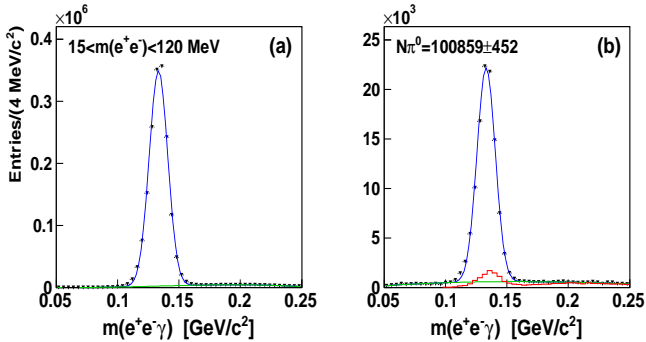}
\caption{ (Color online)
 $m(e^+e^-\gamma)$ invariant-mass distributions obtained in the analysis of Run-I for
 the $m(e^+e^-)$ range from 15 to 120~MeV/$c^2$ with $\gamma p\to e^+e^-\gamma p$ candidates
 selected with the kinematic-fit CL$>$1\%, a $dE/dx$ PID cut accepting the entire
 range with deposits from $e^{+/-}$, and allowing both $e^+$ and $e^-$ to be identified
 with the same PID element:
 (a)~MC simulation of $\gamma p\to \pi^0 p \to e^+e^-\gamma p$ (black dots) fitted with
     the sum of a Gaussian (blue line) for the actual $\pi^0 \to e^+e^-\gamma$ peak
     and a polynomial (green line) of order 4 for the background from misidentifying 
     the recoil proton as either $e^+$ or $e^-$;
 (b)~experimental spectrum (black dots) after subtracting
     the background remaining from $\gamma p\to \pi^0 p \to \gamma\gamma p$.
     The $\pi^0\to \gamma\gamma$ background, which is shown by a red line,
     is normalized to the number of subtracted events.
     The experimental distribution is fitted with the sum of a Gaussian (blue line)
     for the $\pi^0 \to e^+e^-\gamma$ peak and a polynomial (green line) of order 4
     for the background.
}
 \label{fig:eegz34_pi0_cth_2008_fit11_m15_120} 
\end{figure}
\begin{figure}
\includegraphics[width=0.45\textwidth,height=4.35cm,bbllx=1.cm,bblly=.5cm,bburx=19.5cm,bbury=9.cm]{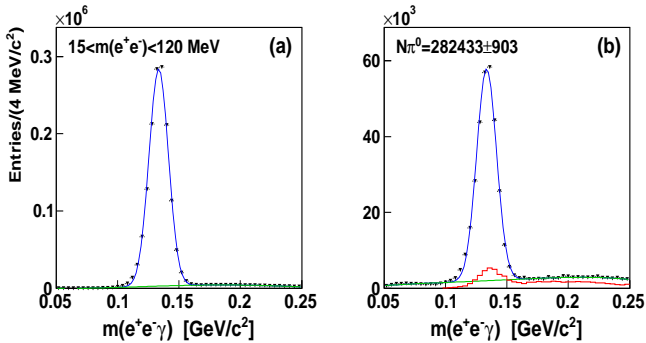}
\caption{ (Color online)
 Same as Fig.~\protect\ref{fig:eegz34_pi0_cth_2008_fit11_m15_120}, but for Run-II.
}
 \label{fig:eegz34_pi0_cth_2013_fit11_m15_120} 
\end{figure}
\begin{figure}
\includegraphics[width=0.45\textwidth,height=4.35cm,bbllx=1.cm,bblly=.5cm,bburx=19.5cm,bbury=9.cm]{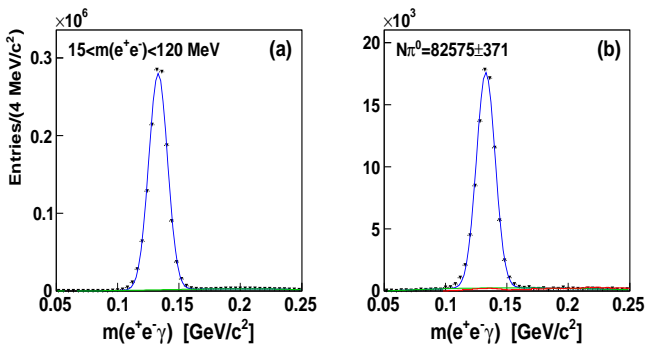}
\caption{ (Color online)
 Same as Fig.~\protect\ref{fig:eegz34_pi0_cth_2008_fit11_m15_120},
 but requiring both $e^+$ and $e^-$ to be identified
 by different PID elements.
}
 \label{fig:eegz34_pi0_cth_2008_fit1_m15_120} 
\end{figure}

 The fitting procedure for $\pi^0 \to e^+e^-\gamma$ and the impact of selection
 criteria on the background is illustrated in
 Figs.~\ref{fig:eegz34_pi0_cth_2008_fit11_m15_120}--\ref{fig:eegz34_pi0_cth_2008_fit1_m15_120}.
 Figure~\ref{fig:eegz34_pi0_cth_2008_fit11_m15_120} shows all
 $\gamma p\to e^+e^-\gamma p$ candidates from Run-I in the $m(e^+e^-)$
 range from 15 to 120~MeV/$c^2$, which were selected with
 the kinematic-fit CL$>$1\%, a $dE/dx$ PID cut accepting the entire range with deposits
 from $e^{+/-}$, and also allowing both $e^+$ and $e^-$ to be identified with
 the same PID element. Figure~\ref{fig:eegz34_pi0_cth_2008_fit11_m15_120}(a)
 depicts the $m(e^+e^-\gamma)$ invariant-mass distribution for the MC simulation
 of $\gamma p\to \pi^0 p \to e^+e^-\gamma p$ fitted with
 the sum of a Gaussian for the actual $\pi^0 \to e^+e^-\gamma$ peak and
 a polynomial of order 4 for the background due to misidentifying 
 the recoil proton as either $e^+$ or $e^-$.
 As shown, the background is very small, especially after the $dE/dx$ PID cut.
 The experimental distribution after subtracting the random and empty-target backgrounds
 and the background remaining from $\gamma p\to \pi^0 p \to \gamma\gamma p$
 is shown by black points in Fig.~\ref{fig:eegz34_pi0_cth_2008_fit11_m15_120}(b).
 The distribution for the $\pi^0 \to \gamma\gamma$ background is normalized
 to the number of subtracted events and is shown in the same figure by
 a red solid line.
 The subtraction normalization was based on the number of events generated
 for $\gamma p\to \pi^0 p \to \gamma\gamma p$ and the number
 of $\gamma p\to \pi^0 p$ events produced in the experiment.
 The experimental distribution was fitted with the sum
 of a Gaussian for the $\pi^0 \to e^+e^-\gamma$ peak and
 a polynomial of order 4 for the background.
 The centroid and width of the Gaussian obtained in both the fits
 (to the MC-simulation and experimental spectra) are in good agreement
 with each other. This confirms the agreement of the experimental data
 and the MC simulation in the energy calibration of the calorimeters
 and their resolution.
 The order of the polynomial was chosen to be sufficient for a reasonable
 description of the background distribution in the range of fitting.
\begin{figure*}
\includegraphics[width=0.85\textwidth]{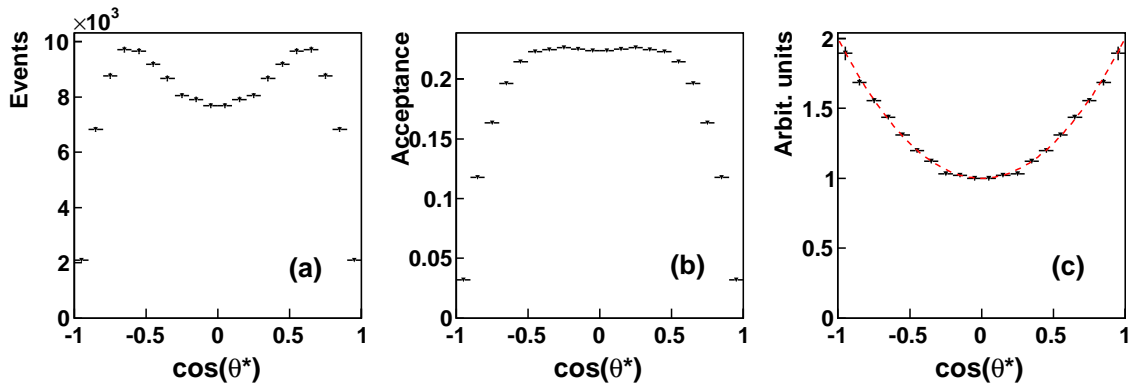}
\caption{ (Color online)
 The $\pi^0\to \gamma\gamma^* \to \gamma e^+e^-$ angular dependence
 (in the $\pi^0$ rest frame)
 of the virtual photon decaying into a $e^+e^-$ pair, with $\theta^*$
 being the angle between the direction of one of the leptons in
 the virtual-photon (or the dilepton) rest frame and the direction
 of the dilepton system (which is opposite to the $\gamma$ direction):
 (a) experimental events from the $\pi^0\to \gamma e^+e^-$ peak;
 (b) angular acceptance based on the MC simulation;
 (c) the experimental spectrum corrected for the acceptance and
     normalized for comparing to the $1 + \cos^2\theta^*$ dependence
     (shown by a red dashed line).
 Because $e^+$ and $e^-$ cannot be separated in the present experiment,
 the angles of both leptons were used, resulting in a symmetric shape
 with respect to $\cos\theta^*=0.$
}
 \label{fig:pi0_eeg_cth2e_a2_2008_3x1} 
\end{figure*}

 The number of $\pi^0 \to e^+e^-\gamma$ decays in both the MC-simulation
 and the experimental $m(e^+e^-\gamma)$ spectra was determined from the area
 under the Gaussian. For the selection criteria and the $m(e^+e^-)$ range
 used to obtain the spectra in Fig.~\ref{fig:eegz34_pi0_cth_2008_fit11_m15_120},
 the averaged detection efficiency was determined to be 23.2\%.

 Figure~\ref{fig:eegz34_pi0_cth_2013_fit11_m15_120} depicts
 the $\pi^0 \to e^+e^-\gamma$ sample obtained from Run-II.
 The selection criteria here were identical to the cuts
 used to plot Fig.~\ref{fig:eegz34_pi0_cth_2008_fit11_m15_120}.
 As shown, the experimental statistic of Run-II is almost
 three times larger, compared to Run-I. However, the PID energy
 resolution was poorer in Run-II, allowing slightly more background
 under the $\pi^0 \to e^+e^-\gamma$ peak and resulting in a slightly
 lower detection efficiency.

 Using events of Run-I, Fig.~\ref{fig:eegz34_pi0_cth_2008_fit1_m15_120}
 illustrates the effect of requiring both $e^+$ and $e^-$ to be identified
 by different PID elements.
 As seen, compared to Fig.~\ref{fig:eegz34_pi0_cth_2008_fit11_m15_120}(b),
 the level of background contributions, including $\pi^0 \to \gamma\gamma$,
 under the $\pi^0 \to e^+e^-\gamma$ peak becomes very small,
 whereas the average detection efficiency decreases to 18.7\%.
 The results for the $\pi^0 \to e^+e^-\gamma$ yield, obtained
 with and without adding events with $e^+$ and $e^-$ identified by
 the same PID element, showed good agreement within the fit uncertainties,
 confirming the reliability in the subtraction of the remaining
 $\pi^0 \to \gamma\gamma$ background.

 The requirement that both $e^+$ and $e^-$ be identified by different PID
 elements results in almost full elimination of the background contributions
 under the $\pi^0 \to e^+e^-\gamma$ peak. This enables measurement of 
 the $\pi^0\to \gamma\gamma^* \to \gamma e^+e^-$ angular dependence
 of the virtual photon decaying into an $ e^+e^-$ pair and comparison
 with Eq.~(\ref{eqn:dtheta}).
 The experimental results for such an angular dependence are
 illustrated in Fig.~\ref{fig:pi0_eeg_cth2e_a2_2008_3x1}
 for events from the $\pi^0 \to e^+e^-\gamma$ peak of Run-I.
 Figure~\ref{fig:pi0_eeg_cth2e_a2_2008_3x1}(a) shows
 the experimental $\cos\theta^*$ distribution.
 The angular acceptance determined from the MC simulation is
 depicted in Fig.~\ref{fig:pi0_eeg_cth2e_a2_2008_3x1}(b).
 The experimental distribution corrected for the acceptance
 is depicted in Fig.~\ref{fig:pi0_eeg_cth2e_a2_2008_3x1}(c) and
 shows good agreement with the expected $1 + \cos^2\theta^*$ dependence.
 The deviation from this dependence due to radiative corrections is just
 few percent at the extreme angles.
 Because $e^+$ and $e^-$ cannot be separated in the present experiment,
 the angles of both leptons were used to measure the dilepton decay
 dependence, which resulted in a symmetric shape with respect to
 $\cos\theta^*=0.$

 The statistics available for Run-I and Run-II and the level of background
 for $\pi^0 \to e^+e^-\gamma$ decays enabled division of all
 candidate events into 18 bins, covering the $m(e^+e^-)$
 range from 15 to 120~MeV/$c^2$. The bins are 5 MeV wide
 up to 90~MeV/$c^2$, and 10 MeV wide at higher masses. Fits to the spectra
 were made separately for Run-I and Run-II, and the final results
 were combined together as independent measurements.
 The fitting procedure was the same as shown in
 Figs.~\ref{fig:eegz34_pi0_cth_2008_fit11_m15_120}--\ref{fig:eegz34_pi0_cth_2008_fit1_m15_120}.

\section{Results and discussion}
  \label{sec:Results}

 The total number of $\pi^0 \to e^+e^-\gamma$
 decays initially produced in each $m(e^+e^-)$ bin was obtained
 by correcting the number of decays observed in each bin
 with the corresponding detection efficiency.
 The results for $|F_{\pi^0\gamma}(m_{e^+e^-})|^2$ were obtained
 from those initial numbers of $\pi^0 \to e^+e^-\gamma$ decays by
 taking into account the total number of $\pi^0\to \gamma\gamma$ decays
 produced in the same data sets~\cite{hornidge13,pi0_a2_2015}
 and the $[{\rm{QED}}(m_{ee})]$ term from Eq.~(\ref{eqn:dgdm_pi0})
 after radiative corrections according to
 the calculations of Ref.~\cite{Husek_2015}.
 The uncertainty in an individual $|F_{\pi^0\gamma}(m_{e^+e^-})|^2$
 value from a particular fit was based on the uncertainty in the number
 of decays determined by this fit (i.e, the uncertainty in the area
 under the Gaussian).
\begin{figure*}
\includegraphics[width=1.\textwidth]{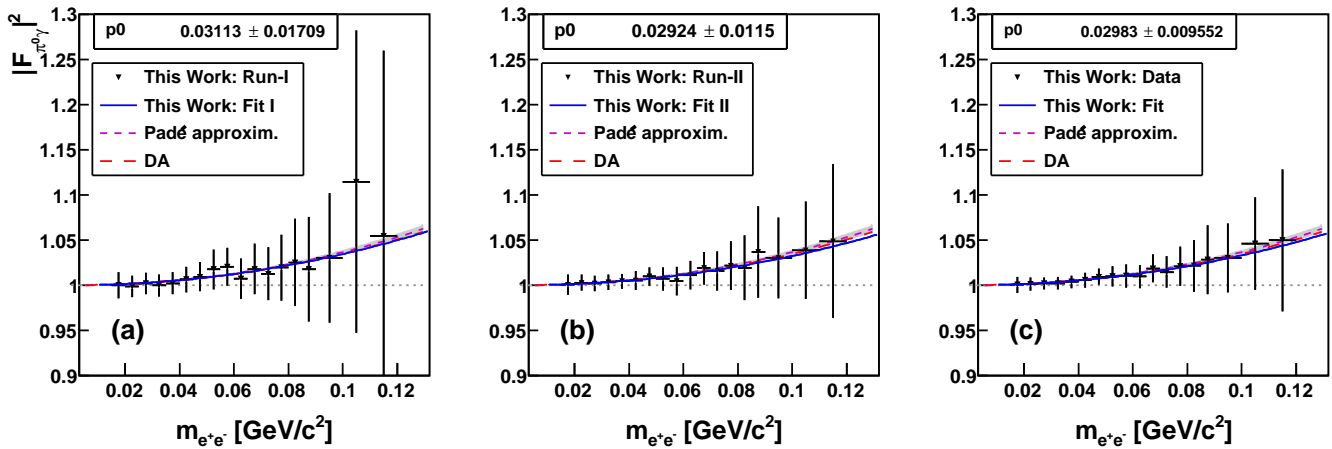}
\caption{ (Color online)
 $|F_{\pi^0\gamma}|^2$ results (black filled triangles) obtained
 from Run-I (a), Run-II (b), and the combined values (c)
 are fitted with Eq.~(\protect\ref{eqn:Fm}) (shown by blue lines,
 with $p0$ being the slope parameter $a_\pi$)
 and compared to the calculations with Pad\'e approximants~\protect\cite{Mas12}
 (shown by a short-dashed magenta line with an error band)
 and to the dispersive analysis (DA) from Ref.~\protect\cite{Hoferichter_2014}
 (long-dashed red line). The error band for the latter analysis is narrower
 by a factor of 4, compared to the other shown, and was omitted because
 of its smallness. The error bars on all data points represent the total
 uncertainties of the results.
}
 \label{fig:tff_pi0_eeg_cth2_radcr_ebst_a2_th_3x1} 
\end{figure*}
 The systematic uncertainties in the $|F_{\pi^0\gamma}(m_{e^+e^-})|^2$ values were
 estimated for each individual $m(e^+e^-)$ bin by repeating its fitting procedure
 several times after refilling the $m(e^+e^-\gamma)$ spectra with different
 combinations of selection criteria, which were used to improve
 the signal-to-background ratio, or after slight changes in the parametrization
 of the background under the signal peak. The changes in selection criteria
 included cuts on the kinematic-fit CL (such as 1\% 2\%, 5\%, and 10\%),
 different cuts on PID $dE/dx$, and switching on and off the requirement
 for both $e^+$ and $e^-$ to be identified by different PID elements.
 The requirement of making several fits for each $m(e^+e^-)$ bin provided a check
 on the stability of the $|F_{\pi^0\gamma}(m_{e^+e^-})|^2$ results.
 The average of the results of all fits made for one bin was then used
 to obtain final TFF values that were more reliable than the results
 based on the fit with the largest number of $\pi^0 \to e^+e^-\gamma$ decays,
 corresponding to the initial selection criteria.
 Because the fits for a given $m(e^+e^-)$ bin with different selection
 criteria or different background parametrizations were based on the same
 initial data sample, the corresponding $|F_{\pi^0\gamma}(m_{e^+e^-})|^2$
 results were correlated and could not be considered as independent measurements
 for calculating the uncertainty in the averaged TFF value.
 Thus, this uncertainty was taken from the fit with the largest number of
 $\pi^0 \to e^+e^-\gamma$ decays in the $m(e^+e^-)$ bin, which was a conservative
 estimate of the uncertainty in the averaged TFF value.
 The systematic uncertainty in the averaged $|F_{\pi^0\gamma}(m_{e^+e^-})|^2$ value
 was taken as the root mean square of the results from all fits
 made for this bin. The total uncertainty in this $|F_{\pi^0\gamma}(m_{e^+e^-})|^2$ value
 was calculated by adding in quadrature its fit (partially reflecting
 experimental statistics in the bin) and systematic uncertainties.
 In the end, the $|F_{\pi^0\gamma}(m_{e^+e^-})|^2$ results from Run-I and Run-II,
 which were independent measurements, were combined as a weighted average
 with weights taken as inverse values of their total uncertainties in quadrature.

 The individual $|F_{\pi^0\gamma}(m_{e^+e^-})|^2$ results obtained from Run-I, Run-II,
 and their weighted average are depicted in Figs.~\ref{fig:tff_pi0_eeg_cth2_radcr_ebst_a2_th_3x1}(a),
 (b), and (c), respectively. The error bars plotted on all data points represent the total
 uncertainties of the results. Fits of the data points with Eq.~(\ref{eqn:Fm})
 are shown by the blue solid lines. The fit parameter $p0$ corresponds to
 the slope parameter $a_\pi$. Because the fits are made to the data points with
 their total uncertainties, the fit errors for $a_\pi$ give their
 total uncertainty as well. Fits that included a normalization parameter
 showed no need for such a parameter, so it was neglected in the end.
 The present experimental results depicted
 in Fig.~\ref{fig:tff_pi0_eeg_cth2_radcr_ebst_a2_th_3x1} are also compared to
 the calculations with Pad\'e approximants~\cite{Mas12} and
 to the dispersive analysis (DA) from Ref.~\cite{Hoferichter_2014},
 which were discussed in the Introduction.
 As shown, all fits to the data points lie slightly lower than the calculations.
 However, the magnitude of the deviation is well within the experimental uncertainties.
 In addition, attempts to fit the present data points with Eq.~(\ref{eqn:Fm2})
 could not provide any reliable values for the curvature parameter $b_\pi$ and
 resulted in a strong correlation between the parameters $a_\pi$ and $b_\pi$.
 The comparison of the individual results obtained from Run-I and Run-II
 illustrates their good consistency within the error bars, even though the uncertainties
 from Run-I are significantly larger than those from Run-II.
\begin{table*}
\caption 
[tab:pi0tff]{
 Results of this work for the $\pi^0$ TFF, $|F_{\pi^0\gamma}|^2$, as a function of
 the invariant mass $m(e^+e^-)$, listed for Run-I, Run-II, and their average,
 where the two uncertainties listed for Run-I and Run-II are fit (reflecting
 statistics) and systematic, respectively, and the total uncertainty is listed
 for the average.
 } \label{tab:pi0tff}
\begin{ruledtabular}
\begin{tabular}{|c|c|c|c|c|} 
\hline
 $m(e^+e^-)$~[MeV/$c^2$]
 & $17.5\pm2.5$ & $22.5\pm2.5$ & $27.5\pm2.5$ & $32.5\pm2.5$ \\
\hline
 Run-I 
 & $1.0001\pm0.0140\pm0.0035$ & $0.9987\pm0.0114\pm0.0033$ & $1.0018\pm0.0110\pm0.0044$ & $0.9996\pm0.0110\pm0.0050$ \\
\hline
 Run-II 
 & $1.0003\pm0.0105\pm0.0036$ & $1.0027\pm0.0085\pm0.0026$ & $1.0019\pm0.0078\pm0.0032$ & $1.0034\pm0.0083\pm0.0020$ \\
\hline
 Run-I + Run-II 
 & $1.0002\pm0.0088$ & $1.0013\pm0.0071$ & $1.0018\pm0.0069$ & $1.0021\pm0.0070$ \\
\hline
\hline
 $m(e^+e^-)$~[MeV/$c^2$]
 & $37.5\pm2.5$ & $42.5\pm2.5$ & $47.5\pm2.5$ & $52.5\pm2.5$  \\
\hline
 Run-I
 & $1.0022\pm0.0119\pm0.0034$ & $1.0063\pm0.0132\pm0.0046$ & $1.0093\pm0.0152\pm0.0053$ & $1.0175\pm0.0165\pm0.0148$ \\
\hline
 Run-II 
 & $1.0044\pm0.0080\pm0.0022$ & $1.0053\pm0.0098\pm0.0034$ & $1.0095\pm0.0097\pm0.0040$ & $1.0069\pm0.0125\pm0.0035$ \\
\hline
 Run-I + Run-II 
 & $1.0037\pm0.0069$ & $1.0057\pm0.0084$ & $1.0094\pm0.0088$ & $1.0096\pm0.0112$ \\
\hline
\hline
 $m(e^+e^-)$~[MeV/$c^2$]
  & $57.5\pm2.5$ & $62.5\pm2.5$ & $67.5\pm2.5$ & $72.5\pm2.5$  \\
\hline
 Run-I 
 & $1.0203\pm0.0200\pm0.0068$ & $1.0073\pm0.0207\pm0.0086$ & $1.0179\pm0.0282\pm0.0021$ & $1.0126\pm0.0289\pm0.0042$ \\
\hline
 Run-II 
 & $1.0046\pm0.0124\pm0.0098$ & $1.0109\pm0.0141\pm0.0069$ & $1.0188\pm0.0169\pm0.0068$ & $1.0154\pm0.0205\pm0.0071$ \\
\hline
 Run-I + Run-II 
 & $1.0102\pm0.0126$ & $1.0097\pm0.0129$ & $1.0185\pm0.0153$ & $1.0144\pm0.0174$ \\
\hline
\hline
 $m(e^+e^-)$~[MeV/$c^2$]
 & $77.5\pm2.5$ & $82.5\pm2.5$ & $87.5\pm2.5$ & $95.0\pm5.0$  \\
\hline
 Run-I 
 & $1.0194\pm0.0358\pm0.0065$  & $1.0251\pm0.0480\pm0.0066$ & $1.0178\pm0.0576\pm0.0076$ & $1.0301\pm0.0694\pm0.0184$ \\
\hline
 Run-II 
 & $1.0214\pm0.0251\pm0.0100$ & $1.0192\pm0.0317\pm0.0165$ & $1.0365\pm0.0478\pm0.0167$ & $1.0303\pm0.0430\pm0.0124$ \\
\hline
 Run-I + Run-II 
 & $1.0207\pm0.0217$ & $1.0213\pm0.0288$ & $1.0284\pm0.0382$ & $1.0302\pm0.0380$ \\
\hline
\hline
 $m(e^+e^-)$~[MeV/$c^2$]
 & $105.0\pm5.0$ & $115.0\pm5.0$ &   \\
\hline
 Run-I 
 & $1.115\pm0.167\pm0.011$ & $1.054\pm0.203\pm0.031$ &  & \\
\hline
 Run-II 
 & $1.039\pm0.053\pm0.007$ & $1.049\pm0.083\pm0.019$ & & \\
\hline
 Run-I + Run-II 
 & $1.046\pm0.051$ & $1.050\pm0.079$ & & \\
\hline
\end{tabular}
\end{ruledtabular}
\end{table*}

 Based on the fit to the data points combined from Run-I and Run-II,
 the magnitude obtained for the slope parameter,
\begin{equation}
 a_\pi = 0.030\pm 0.010_{\mathrm{tot}},
\label{eqn:api_this_work}
\end{equation}
 shows, within the uncertainties, good agreement with
 the RPP value, $a_\pi = 0.032\pm0.004$~\cite{PDG}, and
 with the calculations from Ref.~\cite{Mas12},
 $a_\pi = 0.0324\pm0.0012_{\mathrm{stat}}\pm0.0019_{\mathrm{syst}}$, 
 and Ref.~\cite{Hoferichter_2014}, $a_\pi = 0.0307\pm0.0006$.
 Though the uncertainty obtained for $a_\pi$ in the present measurement
 is significantly larger than in Refs.~\cite{PDG,Mas12,Hoferichter_2014},
 the present result significantly improves the precision in the slope parameter
 $a_\pi$ measured in the timelike region directly from
 the $\pi^0 \to e^+e^-\gamma$ decay and is much
 closer to the precision of the slope parameter extracted from
 the spacelike data~\cite{CELLO_1991}.
 The latest result from NA62, $a_\pi = 0.0368\pm0.0057_{\mathrm{tot}}$~\cite{NA62_2016_final},
 is somewhat greater than all mentioned values but is consistent with them within
 the uncertainties.
 
 The numerical values for the individual $|F_{\pi^0\gamma}(m_{e^+e^-})|^2$ results
 from Run-I and Run-II and for their weighted average
 are listed in Table~\ref{tab:pi0tff}.
 To illustrate the magnitude of each kind of uncertainty, the individual
 results from Run-I and Run-II are listed with both fit and systematic
 uncertainties. The combined results are given with their total uncertainties.
 As shown in Table~\ref{tab:pi0tff}, the total uncertainties
 are dominated by the contribution from the fit uncertainties,
 reflecting statistics. Thus, a more precise measurement
 of the $\pi^0$ TFF at low momentum transfer with the Dalitz
 decay $\pi^0 \to e^+e^-\gamma$ needs a significant increase in experimental statistics.
 The $\pi^0$ TFF parameters extracted from such a precision measurement
 could then constrain calculations that estimate the pion-exchange term, $a_{\mu}^{\pi^0}$,
 to the HLbL scattering contribution to $(g-2)_\mu$.

\section{Summary and conclusions}
\label{sec:Conclusion}
 The Dalitz decay $\pi^0 \to e^+e^-\gamma$ has
 been measured in the $\gamma p\to \pi^0 p$ reaction
 with the A2 tagged-photon facility at the Mainz Microtron, MAMI.
 The value obtained for the slope parameter
 of the $\pi^0$ e/m TFF, $a_\pi=0.030\pm 0.010_{\mathrm{tot}}$,
 agrees within the uncertainties with existing measurements
 of this decay and with recent theoretical calculations.
 The uncertainty obtained in the value of $a_\pi$
 is lower than in previous results based on
 the $\pi^0\to e^+e^-\gamma$ decay.
 The results of this work also include $|F_{\pi^0\gamma}(m_{ee})|^2$ data points
 with their total uncertainties, which allows a more fair comparison of the
 experimental data with theoretical calculations or the use of those data
 in model-independent fits.
 A much more precise measurement of the $\pi^0$ TFF at low momentum transfer
 with the Dalitz decay $\pi^0 \to e^+e^-\gamma$, which has already been planned by the A2
 Collaboration, hopefully will reach the accuracy needed to constrain calculations
 that estimate the pion-exchange term, $a_{\mu}^{\pi^0}$, to the HLbL scattering
 contribution to $(g-2)_\mu$. 

\section*{Acknowledgments}

 The authors wish to acknowledge the excellent support of the accelerator group and
 operators of MAMI.
 We would like to thank Bastian Kubis, Stefan Leupold, and Pere Masjuan
 for useful discussions and continuous interest in the paper.
 This work was supported by the Deutsche Forschungsgemeinschaft (SFB443,
 SFB/TR16, and SFB1044), DFG-RFBR (Grant No. 09-02-91330), the European Community-Research
 Infrastructure Activity under the FP6 ``Structuring the European Research Area''
 program (Hadron Physics, Contract No. RII3-CT-2004-506078), Schweizerischer
 Nationalfonds (Contract Nos. 200020-156983, 132799, 121781, 117601, 113511),
 the U.K. Science and Technology Facilities Council (STFC 57071/1, 50727/1),
the U.S. Department of Energy (Offices of Science and Nuclear Physics,
 Award Nos. DE-FG02-99-ER41110, DE-FG02-88ER40415, DE-FG02-01-ER41194)
 and National Science Foundation (Grant Nos. PHY-1039130, IIA-1358175),
 NSERC of Canada (Grant Nos. 371543-2012, SAPPJ-2015-00023), and INFN (Italy).
 We thank the undergraduate students of Mount Allison University
 and The George Washington University for their assistance.

\end{document}